\documentclass[sigconf]{acmart}

\usepackage{url}
\usepackage{balance}
\usepackage{graphicx}
\usepackage{tabularx}

\begin{document}

\title{Fast Fixes and Faulty Drivers: An Empirical Analysis of Regression Bug
  Fixing Times in the Linux Kernel}

\author{Jukka Ruohonen}
\affiliation{\institution{University of Southern Denmark}
  \city{S{\o}nderborg}\country{Denmark}}
\email{juk@mmmi.sdu.dk}

\author{Adam Alami}
\affiliation{\institution{University of Southern Denmark}
  \city{S{\o}nderborg}\country{Denmark}}
\email{adal@mmmi.sdu.dk}

\begin{abstract}
Regression bugs refer to situations in which something that worked previously no
longer works currently. Such bugs have been pronounced in the Linux kernel. The
paper focuses on regression bug tracking in the kernel by considering the time
required to fix regression bugs. The dataset examined is based on the
\texttt{regzbot} automation framework for tracking regressions in the Linux
kernel. According to the results, (i)~regression bug fixing times have been
faster than previously reported; between 2021 and 2024, on average, it has taken
less than a month to fix regression bugs.  It is further evident that
(ii)~device drivers constitute the most prone subsystem for regression bugs, and
also the fixing times vary across the kernel's subsystems. Although (iii)~most
commits fixing regression bugs have been reviewed, tested, or both, the kernel's
code reviewing and manual testing practices do not explain the fixing
times. Likewise, (iv) there is only a weak signal that code churn might
contribute to explaining the fixing times statistically. Finally, (v) some
subsystems exhibit strong effects for explaining the bug fixing times
statistically, although overall statistical performance is modest but not
atypical to the research domain. With these empirical results, the paper
contributes to the efforts to better understand software regressions and their
tracking in the Linux kernel.
\end{abstract}

%



\maketitle

\section{Introduction}

A regression bug, or simply a regression, is a bug related to a feature that
does not currently work, although it worked previously. It is possible to
consider regressions at different levels of abstraction, including a
fine-grained level of commits to a version control system. When operating at a
higher level of software releases, a regression is a bug that exists in some
version but does not exist in a previous version or a software's whole previous
version history.

To mitigate regression bugs, a simple rule of thumb is often used: any
regression bug fix should come with a test case that can later on verify that
the bug will not reoccur in the future. In other words, regression ``testing
involves repetitive tests and aims to verify that previously working software
still works after changes to other parts'' \cite[p.~10]{Engstrom10}. In addition
to testing, code reviews are often used too as a mitigative
measure~\citep{Braz22}. Nevertheless, test suites that can be automatically
executed align well with so-called continuous software engineering practices,
including continuous integration and continuous delivery in
particular~\citep{Stolberg09}. These practices, in turn, align with other
current software development paradigms, such as continuous
refactoring. Refactoring is important also in the highly complex context of
operating system kernels for paying back technical and architectural
debt~\cite{Spinellis21}. However, it should be emphasized that refactoring can
also cause regression bugs~\citep{Rachatasumrit12}. Thus, automated tests are
important also for improving refactoring.

Refactoring is inevitably related also to the concept of code churn, which
generally refers to rapid and recurrent commit-by-commit changes made to a
version control system. Due to many new features constantly introduced,
deprecating of old features, constant refactoring, bug fixes, and a large amount
of developers, the Linux kernel is also notable for its extensive amount of code
churn~\citep{Palix11, Ruohonen19RSDA}. These characteristics of the Linux kernel
development help to understand why also regression bugs have been very
common. In fact, it has been reported that as much as a half of all bugs in the
Linux kernel have been regressions, and new features or changes to existing
features account for over half of the regression bugs~\citep{Xiao19}. New
features are also the most typical reason behind vulnerabilities
introduced~\citep{JiangJiang24}. The present paper continues this line of
previous empirical research by focusing on regression bug tracking, including
particularly the time required to fix regression~bugs.

The previous results noted justify the paper's relevance; regression bugs
continue to be a big problem in the Linux kernel and its
development~\cite{Xiao19}, and thus shedding more light on these bugs and their
tracking constitute a relevant research topic. Furthermore, this topic has not
been examined previously---at least to the extent presented in the current
work. The Linux kernel is also an always excellent case study due do its
importance, including both in the open source software world and the commercial
software industry.

By regression bug tracking, it is meant that fixing and keeping track of
regression bugs require distinct software engineering activities. In particular,
the starting point is a so-called bisection, a binary search, for finding a
particular commit that introduced a given regression. After finding the commit,
further search may be required to trace follow-up commits that possibly also
contributed to the given regression bug. After these tasks, traditional
so-called bug triaging follows; a developer responsible and available needs to
be located, a given regression bug needs to be evaluated for its severity and
impact, a bug fix needs to be developed, the fix needs to be preferably reviewed
and tested, the fix needs to be integrated into a mainline branch and a future
release, the given bug report needs to be closed in a bug tracking system, and
so forth.

In addition to these rather conventional software engineering tasks, in
large-scale software projects, such as the Linux kernel, it is further important
to systematically archive regression bugs for later analysis, including during
fixing of further regression bugs. For instance, systematic tracking and
archiving are important because sometimes fixing a regression bug introduces a
new regression or even a vulnerability~\citep{JiangJiang24, Xiao20}. A further
point about the importance of tracking and archiving is that some regression
bugs have security consequences~\citep{Braz22}, which further enlarge the scope
of triaging and coordination, and the need for traceability. Archiving provides
also data for software analytics. Due to these and related reasons,
\texttt{regzbot}, an automation solution for regression bug tracking in the
Linux kernel was introduced~\citep{regzbot24a}. The present paper analyzes data
extracted from the automation bot, thus also demonstrating its value for
software analytics and empirical software engineering research. It can be also
remarked that \texttt{regzbot} provides a small counterargument to recently
expressed criticism that bug detection and automation tools in general have
still received limited use in the Linux kernel
development~\citep{Suzuki24}. Against this backdrop, it should be also
emphasized that the Linux kernel development relies heavily on code reviews for
improving software quality. While code reviews should supposedly reduce the
amount of regression bugs, peer reviewing might also decrease regression bug
fixing times because multiple people have already been involved. Once a new
regression bug is reported, it should be easy to find also these other
developers who are already familiar with the contributing faulty commit due to
their reviewing activity.

With these motivating points in mind, the present paper examines the following
five research questions (RQs):
\begin{itemize}
\itemsep 3pt
\item{RQ.1: How long does it take to fix regression bugs in the Linux kernel?}
\item{RQ.2: Which kernel subsystems are prone to regression bugs, and do the bug
  fixing times vary across subsystems?}
\item{RQ.3: How many of the commits that fixed regressions were reviewed,
  tested, or both reviewed and tested, and do the associated regression bug fixing times vary in terms of reviewing, testing, or both?}
\item{RQ.4: Can code churn explain the bug fixing times?}
\item{RQ.5: How well the regression bug fixing times can be predicted, and what
  factors particularly explain the fixing times?}
\end{itemize}

The five research questions are further motivated in the opening
Section~\ref{sec: related work} by better connecting them to existing
research. Then, Section~\ref{sec: data and methods} elaborates the empirical
data used and the methodology for analyzing it, including the software metrics
derived and used for answering to RQ.5. Results are presented in the subsequent
Section~\ref{sec: results}. The final Section~\ref{sec: discussion} provides a
concluding discussion.

\section{Related Work}\label{sec: related work}

There are three large branches of related work. The first is the rather
extensive research of the Linux kernel itself. In terms of empirical software
engineering, the myriad of topics examined in this branch range from
architecture~\citep{Bowman99}, evolution~\citep{Israeli10, Passos13}, code
reviews~\citep{Bettenburg15, Rigby11}, and development
processes~\citep{TanZhou20} to bugs~\citep{Ahmed09, Palix11, Suzuki24} and
vulnerabilities~\text{\citep{ChenMao11, JiangJiang24, Jiminez16}}, including
those found via fuzzing~\citep{Kim20, Ruohonen19RSDA}. Although there is also
some existing research on regressions~\citep{Xiao20, Xiao19}, including
performance regressions~\citep{Harji11}, this literature is very limited
compared to the other research topics mentioned. Having said that, when the
scope is extended toward so-called gray literature, also the amount of
explicitly related previous works extend.

Several presentations and associated publications on regressions have made by
practitioners. As could be expected, most of these have focused on a practical
question of how the situation with regressions could be improved. Although
improving software testing is the obvious and classical solution, also other
automation techniques have long been proposed, including with respect to a
problem of identifying a particular commit that introduced a given
regression~\citep{Johnson04, Yoshioka07}. Several articles have also appeared in
open source magazines and professional outlets. In fact, also the developer of
\texttt{regzbot} was recently interviewed in such a
magazine~\citep{Corbet24a}. In the interview he and other Linux kernel
developers regretted the still sorry state of affairs, pointing out limitations
of using the kernel's bug tracking system and mailing lists for triaging and
generally dealing with regression bugs, among other things.

The second branch is again the rather extensive literature on regression
testing. To again pinpoint some areas of research, the topics examined include,
but are not limited to, validation and verification~\citep{Pastore15}, test case
selection and prioritization~\citep{Wong97},
refactoring~\citep{Rachatasumrit12}, challenges faced by practitioners,
including those related to automation, failure identification in test suites,
and cost analysis~~\citep{Engstrom10, Onoma98}, regression testing of graphical
user interfaces~\citep{Menon03}, evaluation of how much a project's past can
regression tests cover~\citep{MaesBermejo24}, and handling of regression bugs
that are security vulnerabilities~\citep{Braz22}. When combining the two
branches of research, it can be stated that, with some notable
exceptions~\citep{Xiao20, Xiao19}, rather limited attention has been devoted to
regression bugs in the Linux \text{kernel---even} though the Linux kernel is of
particular importance in the open source domain and regressions affecting it
have wide consequences for users and deployments. Another point is that there
seems to be only a \text{little---if any---research} on the handling, tracking,
triaging, and archiving of regression bugs. Thus, the five research questions
postulated and the answers given to them have noteworthy novelty.

The third branch of related work is the empirical research on bug fixing
times. Although regression bug fixing times have not been previously examined
according to a reasonable but still non-systematic literature search, the branch
has essentially focused on answering to a similar question than RQ.5. Typical
explanatory factors considered in the research branch include persons who file
bug reports and their characteristics, code churn, amount of commits required to
fix bugs, a volume of discussions on bug reports, amount of online references to
external sources, and a severity of bugs~\citep{Marks11, Perelta24,
  Ruohonen24SANERb, Vieira20}. Regarding the Linux kernel, existing results
indicate that many of the bugs affecting the kernel have traditionally been
rather long-lived, taking about two years to fix on average~\citep{Palix11}. By
following existing research~\citep{Ruohonen19RSDA}, a particular interest in the
present work is to also examine whether commits that fixed regression bugs were
reviewed or tested by other developers~(RQ.3). This question is important due to
the stringent peer review practice endorsed and used in the Linux kernel
development~\cite{Alami22}. The question can be further motivated by existing
studies indicating that vulnerability-introducing commits are reviewed less
often compared to other commits~\citep{JiangJiang24}. By hypothesis, a similar
effect is present with regression bugs, although only fixing of these is
considered in the present work due to limitations in \texttt{regzbot}. As was
already discussed, code churn may well be behind introducing of regression bugs,
and it may also be that it can explain the later fixing times of these
bugs. Although RQ.4 is on the exploratory side of things, the question is still
worth asking and examining.

A further important point is that bugs in operating system kernels typically
vary across different subsystems. Device drivers are a good example in this
regard. With or without scaling by the size of subsystems, device drivers have
indeed been the most bug-prone subsystem in the Linux kernel~\citep{Ahmed09,
  Xiao19, Zhang21}. Although the situation may or may not be similar with
vulnerabilities~\citep{JiangJiang24, Jiminez16, ShameliSendi21}, also many
regression bugs have affected device drivers.

While low-level and often undocumented hardware-dependent details may provide a
partial explanation, another partial explanation may relate to testing
obstacles. Namely, the Linux kernel supports many legacy devices and many niche
devices, both of which may suffer from a lack of testing already because it may
be difficult to find people to test modifications to drivers for such devices
with actual hardware. A third partial explanation may relate to development
practices; some device drivers in the Linux kernel have \text{been---and}
\text{are---written} by employees working in companies producing the associated
hardware. Again, the partial reliance on inside knowledge may limit testing and
lead to maintenance problems in case a given company-specific device driver
needs to be maintained by someone else in the future. The fourth potential
explanation is closely related: each subsystem in the Linux kernel has a
specific subsystem maintainer, and thus it may be that personal preferences,
social interactions, and related factors contribute to variance of bugs,
including regression bugs, across different subsystems. Because kernel
subsystems also connote with subcultures~\citep{Kudrjavets22a}, some subsystems
may emphasize code reviews and manual testing more than others, among other
things. A final potential explanation may be simple: as the Linux kernel
evolution, including its growing size, has often been seen as primarily being
driven by the addition of new drivers~\cite{Passos13, ZhouCheng17}, the
freshness of device driver code may also explain its proneness to bugs. Whatever
the actual explanation might be, the rationale for asking RQ.2 is
well-justified.

Finally, there is RQ.1 to be briefly discussed. It has been answered to in
previous research based on data from the Linux kernel's bug tracking
system~\citep{Xiao19}. To this end, answering to RQ.1 leans toward a replication
check; it is important to know whether the answers remain similar when using
different data sources. It may be that \texttt{regzbot} is a more robust source
than trying to curate regression bugs from the bug tracking system. In any case,
at least it is much more efficient because someone else has already done the
curating and the bot is also used in production for tracking~regressions.

\section{Data and Methods}\label{sec: data and methods}

\subsection{Data}\label{subsec: data}

The data was collected from the \texttt{regzbot}'s online
dashboard~\citep{regzbot24a} in 30 October, 2024. A replication package is
available online for the pre-processed data and the associated statistical
computation routines~\cite{Ruohonen24rep}. Only resolved regression bugs were
included in the dataset. This choice is necessary for an operationalization of a
bug fixing time, which is defined as a time difference between an initial bug
report and the last commit that fixed the given bug according to the tracking
bot. In this regard, it can be mentioned that the bot has a notable limitation
in that it does not do bisection for finding out a given commit that introduced
a given regression. If such functionality would be available, it would be
possible to further extend the analysis toward whole life cycles of regression
bugs.

Also another important point should be raised about the operationalization of
regression bug fixing times: multiple individuals may have reported the same
regression bug, possibly on different reporting venues, such as the kernel's bug
tracking system and its numerous mailing lists. A simple decision was taken to
address this problem: an initial bug report was always taken to refer to the
first report enumerated by \texttt{regzbot}. In theory, it might be possible to
consider all reports, picking the one with the earliest date as the initial
report. In practice, however, the numerous distinct venues make parsing
difficult because there is no uniform format across the venues, some of which
point to external tracking systems used by other open source software
projects. Having said that, quantified information about the other reports and
reporters is still used as an explanatory factor, as soon discussed in the next
section.

\begin{figure}[th!b]
\centering
\includegraphics[width=\linewidth, height=3.5cm]{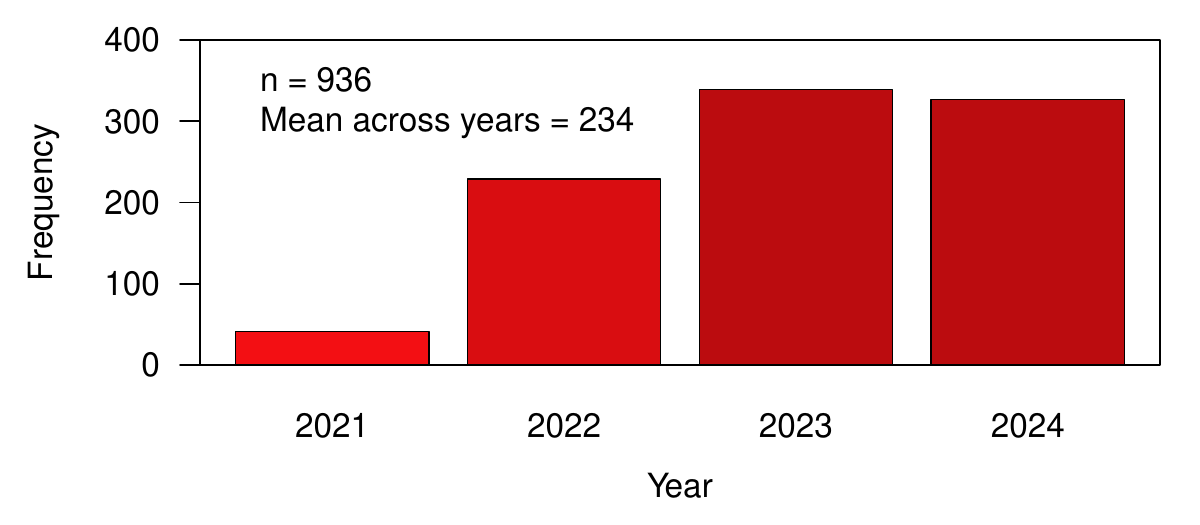}
\caption{The Annual Amounts of Regression Bugs According to the Dates of Initial
  Bug Reports}
\label{fig: years}
\end{figure}

In total, $n = 936$ regression bugs are covered in the dataset. The amount of
observations is lower than in a comparable previous work that has operated with
$n = 2000$ regression bugs~\citep{Xiao19}. The explanation traces to
\texttt{regzbot}, which only covers the kernel's recent past. In fact, as can be
seen from Fig.~\ref{fig: years}, only four years are covered. As only a limited
amount of regression bugs was tracked by the bot in 2021, it seems sensible to
include controls for the annual amounts in the forthcoming statistical and
machine learning modeling.

\subsection{Metrics}\label{subsec: metrics}

Twelve metrics are used to model and predict the regression bug fixing
times. These are enumerated and elaborated in Table~\ref{tab: metrics}. As is
perhaps typical to research operating with software metrics, it is somewhat
difficult to raise a level of abstraction by theorizing what the metrics might
proxy as empirical stand-ins for higher level theoretical
concepts~\cite[cf.][]{Stol24}. Nevertheless, some of these are clearly related
to software development practices and processes. The code reviewing practices in
the Linux kernel are a good example in this regard. Then, the ENTROPY metric
might be taken to proxy interpretability of bug reports, SUBSYSTEM is related to
software architectures, BUGTRACKER might proxy the coordination and triaging
aspects related to regression bugs, and REPORTERS could be seen to partially
proxy the famous ``Linus law'', that is, the notion that ``given enough
eyeballs, all bugs are shallow''~\cite{Raymond01}. These remarks
notwithstanding, a prior and explicit theorization is omitted, although
theoretical implications are revisited in the concluding Section~\ref{sec:
  discussion}. Sketching theoretical premises is also easier once it is known
which particular metrics are strong predictors.

\begin{table}[th!b]
\centering
\caption{Explanatory Metrics}
\label{tab: metrics}
\begin{tabularx}{\linewidth}{lX}
\toprule
Metric & Description \\
\hline
REPORTERS & A number of unique individuals who reported a regression bug. \\
\cmidrule{2-2}
BUGTRACKER & A dummy variable scoring one in case a given regression bug was reported also in the kernel's bug tracking system and not only on any of mailing lists or other reporting venues. \\
\cmidrule{2-2}
REFERENCES & A number of references to emails that \texttt{regzbot} has classified as being related to a given regression bug; see \cite{regzbot24b} for details. The counting includes also the final commit that finally fixed the given regression bug. \\
\cmidrule{2-2}
ENTROPY & Shannon's entropy of an initial bug report. \\
\cmidrule{2-2}
REVTEST & A dummy variable scoring one in case a commit that fixed a given regression bug was reviewed, tested, or both reviewed and tested by at least one other developer besides a given committer. The operationalization is based on searching the \texttt{reviewed-by}, \texttt{tested-by}, and \texttt{reviewed-and-tested-by} character strings from \texttt{git}'s lower-cased commit messages. \\
\cmidrule{2-2}
ACKED & A dummy variable scoring one in case a commit that fixed a bug was acknowledged by someone, as captured by the \texttt{acked-by} string in lower-cased commit message. \\
\cmidrule{2-2} SUGGESTED & A dummy variable scoring one in case a commit that fixed
a bug was suggested by someone, as captured by the \texttt{suggested-by} string
in lower-cased commit message. \\
\cmidrule{2-2}
MODFILE & A number of files modified in a given commit that introduced a particular regression bug. \\
\cmidrule{2-2}
LOCADD & A number of lines of code added in a commit that introduced a given regression bug. \\
\cmidrule{2-2}
LOCDEL & A number of lines of code deleted in a commit that introduced a given regression but. \\
\cmidrule{2-2}
SUBSYSTEM & A set of dummy variables for the subsystems in which the regression bugs observed were located; the reference category is device drivers. See the body of the text for further details. \\
\cmidrule{2-2}
YEAR & Three dummy variables capturing the years in which the bugs were opened
(cf.~Fig.~\ref{fig: years}). The reference category is the year 2021. \\
\hline
\end{tabularx}
\end{table}

A couple of remarks should be made about the operationalization behind the
metrics. First, many of the metrics have been used in previous research on which
also the operationalization is largely based. For instance, like in previous
studies~\citep{Kudrjavets22b, Ruohonen19RSDA}, the three churn metrics, MODFILE,
LOCADD, and LOCDEL, are based on the \texttt{git}'s \texttt{--shortstat} command
line option for the commits fixing regression bugs. Also the operationalization
of REVTEST and ENTROPY is similar than in previous studies~\citep{JiangJiang24,
  Ruohonen18IST, Ruohonen19RSDA}. Second, the operationalization of SUBSYTEM
requires a brief remark too. By again following existing
studies~\cite{JiangJiang24}, the top-level directories are used to proxy
subsystems. In case a commit touched multiple subsystems, the one with the
highest frequency of changes is used, and if there is a tie, the first
subsystem, as listed in a commit message, is~used. A group of ``others'' is used
for directories and files that do not represent subsystems. The examples include
build scripts, header files for C, certificates, and internal kernel documentation.

\subsection{Methods}\label{subsec: methods}

The research branch examining bug fixing times provides an overall guidance for
methodological choices. In essence, the branch has operated with two approaches
to empirical modeling: regression analysis and
classification~\cite{Ruohonen24SANERb}. Both are used in the present work. In
terms of regression analysis, the conventional negative binomial regression is
used, given that the fixing times modeled are a count data variable. Computation
is done with the \texttt{glm.nb} function from the \texttt{MASS}
package~\cite{MASS} for R. By following existing research~\cite{Ruohonen18IST},
the classification is based on a median split; a ``fast group'' is below the
median of the regression bug fixing times and a ``slow group'' has fixing times
equal to the median or greater than it. In terms of computation, the
\texttt{caret}~\cite{caret} package for R is used with a na\"ive
Bayes~\cite{naivebayes}, a~random forest~\cite{rf}, and a support vector machine
(with a linear kernel)~\cite{kernlab} classifiers. Default settings in the
associated packages are used for all computations. A $5$-fold cross-validation
is used for training the classifiers, and a random sample containing ten percent
of all observations in the dataset is used as a test set.

Regarding RQ.5, the phrasing ``how well regression bug fixing times can be
predicted'' requires general purpose performance metrics. With respect to the
regression analysis pursued, numerous so-called pseudo-R$^2$ measures have been
introduced and used over the years~\cite{Hoetker07}. Although none of these
fully resemble the conventional coefficient of determination, $R^2$, in an
ordinary least squares regression, they are useful for heuristic purposes and
for improving comparability of results across studies. To this end, the
well-known pseudo-$R^2$ measure of Nagelkerke \cite{Nagelkerke91} is used. In
addition, the Schwarz's \cite{Schwarz78} well-known Bayesian information
criterion (BIC) is used to compare performance across the regression models
estimated. The choice of performance metrics for the classification approach is
easier: the conventional group of four---precision, recall, accuracy, and the
F1-measure---is used and reported. All are well-known metrics, and thus require
no particular further elaboration.

Then, again regarding RQ.5, the phrasing ``what factors particularly explain the
fixing times'' requires a clarification too. As usual, examining regression
coefficients provides a straightforward way to approach the phrasing. Two
additional modeling technique are used in order to seek a more parsimonious
model than a one including all the twelve metrics. First, both the negative
binomial regression approach and the classification approach are computed by
estimating twelve models for both consecutively by cumulatively adding the
metrics in Table~\ref{tab: metrics}. In other words, in addition to intercepts,
the first model includes only REPORTERS, the second model includes REPORTERS and
BUGTRACKER, and so forth and so on. The models are then evaluated by examining
the performance metrics outlined. Second, with respect to the regression
approach, the \texttt{bestglm} package~\cite{bestglm} for R is used to briefly
examine a result from a more sophisticated and automated model selection
algorithm~\cite{Morgan72} based on BICs. As model selection based on BICs is
known to lead to excessively simple models in applied
research~\cite{Weakliem99}, the result should be interpreted with a grain of
salt, comparing it to the manually computed consecutive estimation results.

\section{Results}\label{sec: results}

\subsection{Fixing Times}\label{subsec: fixing times}

\begin{figure}[th!b]
\centering
\includegraphics[width=\linewidth, height=4.5cm]{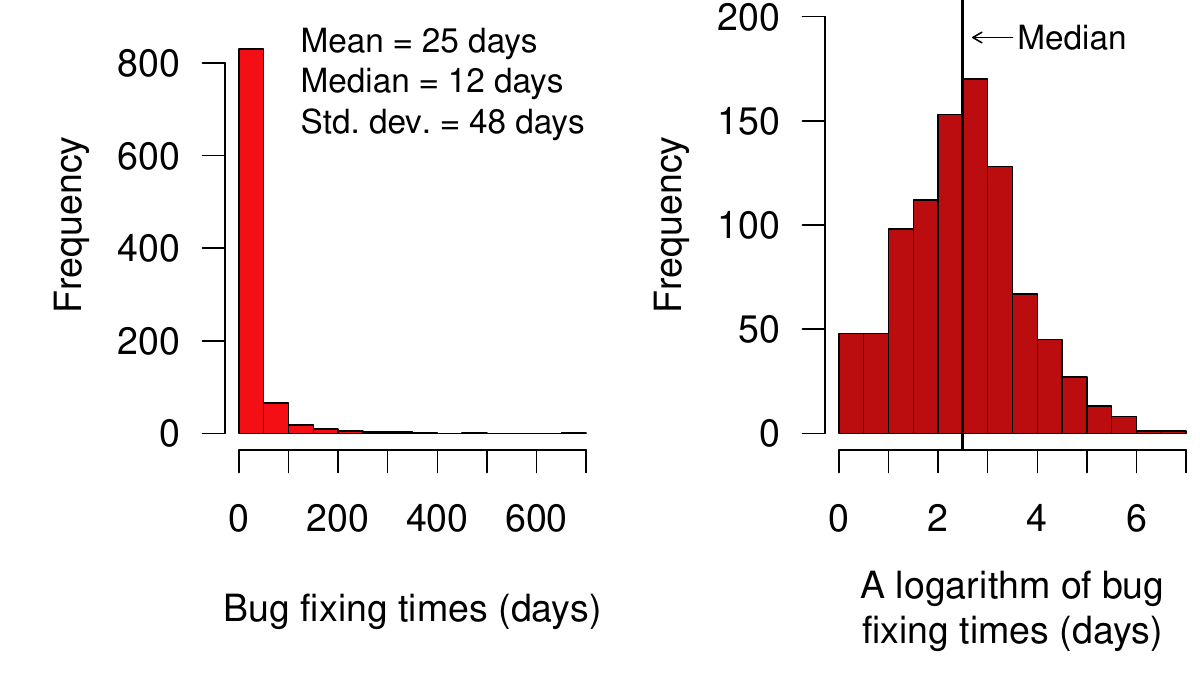}
\caption{Fixing Times}
\label{fig: diff time}
\end{figure}

The presentation of the empirical results is reasonably done by going through
the four research questions consecutively. Thus, Fig.~\ref{fig: diff time} shows
the regression bug fixing times. Although the fixing times exhibit a very
long-tailed distribution, which is typical in the research
branch~\cite{Ruohonen18IST, Ruohonen24SANERb}, the averages indicate quite fast
fixing in general. The arithmetic mean is $25$ days and the median is only $12$
days. These are much lower values than what has been reported previously for the
Linux kernel; an average fixing time of about $109$ days for regression bugs and
$215$ days for non-regression bugs~\cite{Xiao19}. In line with the remarks in
Section~\ref{subsec: data}, a potential explanation may relate to different time
frames, meaning that the fixing might have become faster in recent years. In any
case, the divergence strengthens the replication motivation that was discussed in Section~\ref{sec: related work}.

\subsection{Subsystems}\label{subsec: subsystems}

\begin{figure}[th!b]
\centering
\includegraphics[width=\linewidth, height=4.5cm]{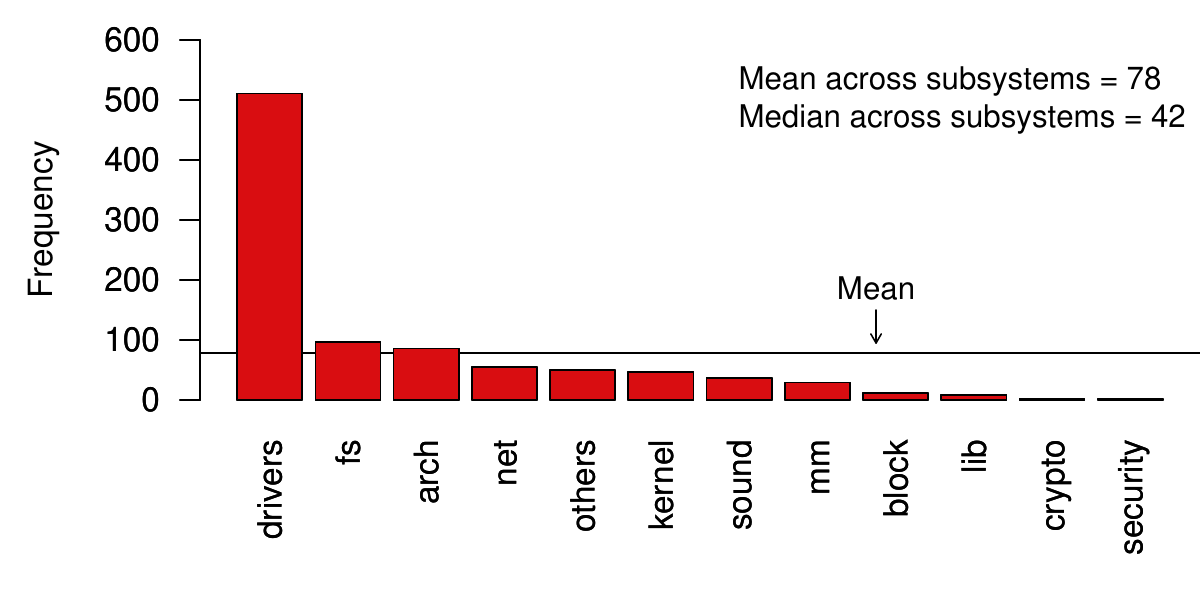}
\caption{Regression Bugs Across Subsystems}
\label{fig: subsystems}
\end{figure}

\begin{figure}[th!b]
\centering
\includegraphics[width=\linewidth, height=4cm]{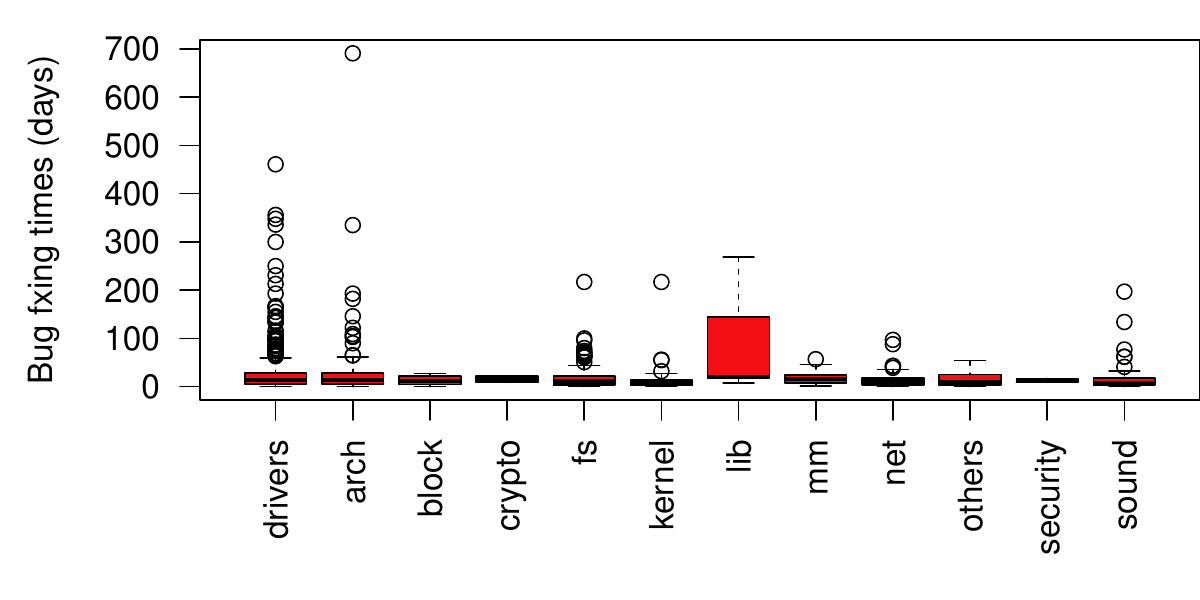}
\caption{Bug Fixing Times Across Subsystems}
\label{fig: subsystems diff}
\end{figure}

The second research question asked about the distribution of regression bugs
across the Linux kernel subsystems, and whether the fixing times also vary
across the subsystems. The answers can be summarized in the form of
Fig.~\ref{fig: subsystems} and Fig.~\ref{fig: subsystems diff}. As can be seen
from the former figure, the conventional wisdom from existing studies holds
well: device drivers have clearly been the most prone subsystem to regression
bugs. In fact, these account for about 55\% of all regression bugs in the
sample. This share is very close to previously reported results that were noted
in the preceding sections.

Also the regression bug fixing times vary across the subsystems. The subsystem
for auxiliary kernel libraries, \texttt{lib}, stands out in this regard. It is
also noteworthy that \texttt{drivers} and \texttt{arch} contain many
outliers. The latter subsystem is about machine-dependent kernel code for
hardware with different instruction set infrastructures. In preparation for the
more formal statistical analysis, it can be also remarked that the classical
Kruskal-Wallis test~\cite{Kruskal52} rejects a null hypothesis that the medians
would be equal across the subsystems.

\subsection{Reviewing and Testing}

The third research question was postulated similarly to RQ.2. In addition to the
REVTEST metric, Table~\ref{tab: binary valued} includes also the other two
metrics with a dichotomous scale. On one hand, it can be concluded that the
majority (60\%) of the commits fixing regression bugs were reviewed, tested, or
both reviewed and tested by developers other than given committers. The values
are generally much higher than what has been reported in a bug fixing
context~\cite{Ruohonen19RSDA}. On the other hand, only a few commits were
further acknowledged by other developers, and even a fewer commits were
suggested by others.

\begin{table}[t!]
\centering
\caption{Bugs Across Three Binary-Valued Metrics}
\label{tab: binary valued}
\begin{tabular}{lrrlrrlrr}
\toprule
& \multicolumn{2}{c}{REVTEST}
&& \multicolumn{2}{c}{ACKED}
&& \multicolumn{2}{c}{SUGGESTED} \\
\cmidrule{2-3}\cmidrule{5-6}\cmidrule{8-9}
& No & Yes && No & Yes && No & Yes \\
\hline
Frequency & 378 & 558 && 738 & 138 && 884 & 52 \\
Share (\%) & 40 & 60 && 85 & 15 && 94 & 6 \\
\hline
\end{tabular}
\end{table}

\begin{figure}[t!]
\centering
\includegraphics[width=\linewidth, height=4cm]{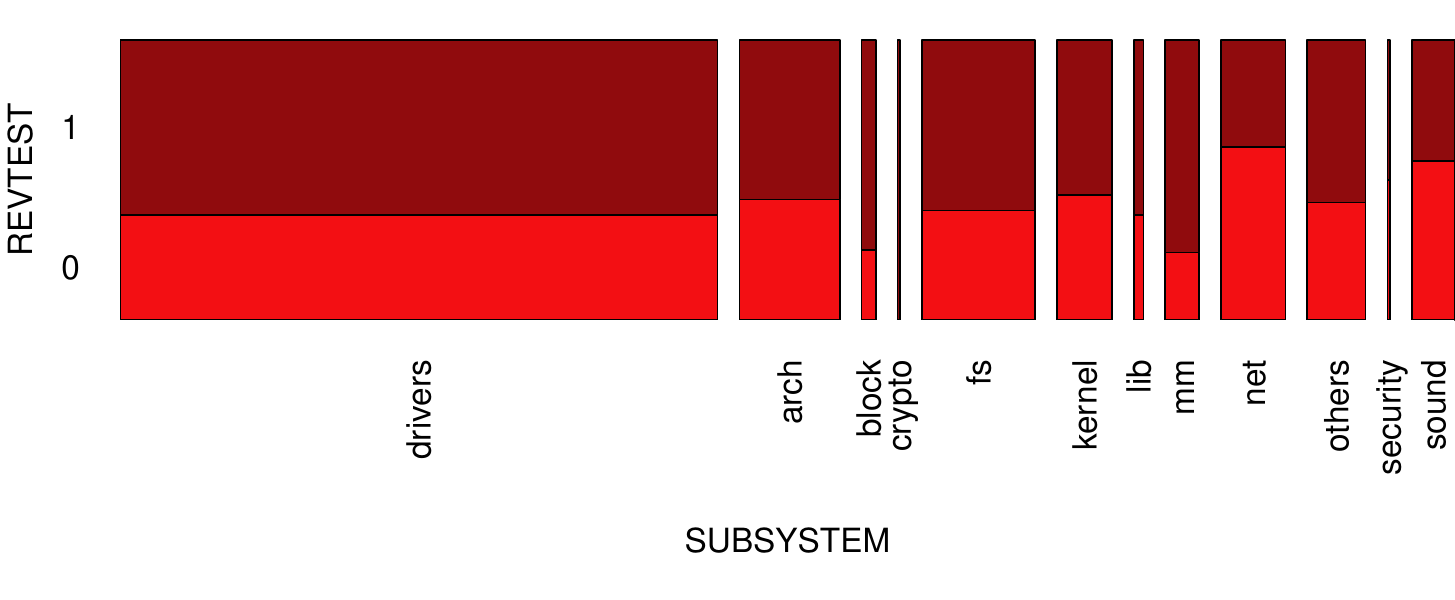}
\caption{Testing and Reviewing Across Subsystems}
\label{fig: revtest subsystem}
\end{figure}

\begin{table}[t!]
\centering
\caption{Correlations (Pearson) Between Regression Bug Fixing Times and Code Churn Metrics}
\label{tab: code churn cor}
\begin{tabular}{rrlrrlrr}
\toprule
\multicolumn{2}{c}{MODFILE}
&& \multicolumn{2}{c}{LOCADD}
&& \multicolumn{2}{c}{LOCDEL} \\
\cmidrule{1-2}\cmidrule{4-5}\cmidrule{7-8}
Coef. & $p$-value && Coef. & $p$-value && Coef. & $p$-value \\
\hline
0.03 & 0.36 && \textbf{0.18} & <0.01 && 0.04 & 0.21 \\
\hline
\multicolumn{6}{l}{~~\small{Note: boldfaced coefficients denote $p < 0.05$.}}
\end{tabular}
\end{table}

The second part of RQ.3 can be again briefly summarized with Kruskal-Wallis
tests, which do not reject the null hypotheses for all three metrics (with
$p$-values of $0.69$, $0.15$, and $0.40$ for REVTEST, ACKED, and SUGGESTED,
respectively). Thus, it can be already at this point concluded tentatively that
reviewing and testing do not increase (or decrease) the regression bug fixing
times---even though both software engineering activities take additional time;
historically, on average it has taken less than ten days to complete peer
reviews in the Linux kernel~\cite{Bettenburg15}. If the line of thought in
Section~\ref{subsec: metrics} is continued a little by linking ACKED and
SUGGESTED to auxiliary coordination in regression bug fixing, neither such
coordination seems to affect the regression bug fixing times in the Linux
kernel. However, it should be noted that also reviewing and testing vary a
little across the kernel's subsystems. As can be seen from Fig.~\ref{fig:
  revtest subsystem}, the \texttt{net} subsystem is an outlier because within
there reviewing and testing have been lower than in other subsystems~on~average.

\subsection{Code Churn}

The fourth research question contemplated an exploratory hypothesis that code
churn might partially explain the regression bug fixing times in the Linux
kernel. Although the later statistical analysis provides a more robust
conclusion, a preliminary answer can be motivated by looking at the correlations
shown in Table~\ref{tab: code churn cor}.

Of the three churn metrics, only LOCADD, a number of lines added in commits
fixing regression bugs, is statistically significant at the conventional $p <
0.05$ threshold. However, the Pearson's product moment correlation coefficient
is small in magnitude. By examining a scatter plot between the two metrics, it
becomes also evident that the weak linear relation is almost entirely explained
by a few extreme outliers in LOCADD. With these points in mind, it can be
tentatively concluded that the answer to RQ.4 is negative.

\subsection{Statistical Modeling and Predictions}

The more formal statistical modeling can be started by examining results from
the negative binomial regression analysis. Thus, the coefficients from a full
model including all available explanatory metrics are shown in Table~\ref{tab:
  full model}. Of the $25$ parameters in the model, in addition to the
intercept, only eight are statistically significant at the zero point zero five
level. Among these is also LOCADD. However, analogously to the previous
correlation results in Table~\ref{tab: code churn cor}, the magnitude of the
coefficient for the metric is very small. A~similar point applies to REFERENCES,
another metric with a continuous scale. In contrast, some of the SUBSYSTEM dummy
variables have relatively large magnitudes. When compared to \texttt{drivers},
the \texttt{lib} subsystem has had slower fixing times, as could be also
expected from the earlier discussion in Section~\ref{subsec: subsystems}. Also
\texttt{arch} has had slower fixing times than the device driver subsystem, but
this result might be partially explained by the single extreme outlier visible
in Fig.~\ref{fig: subsystems diff}. The remaining statistically significant
SUBSYSTEM dummy variables have negative signs, indicating faster fixing times
than regression bugs in the \texttt{drivers} subsystem. Finally, the dummy
variable for the year 2024 is also statistically significant. This observation
provides empirical support for the point earlier raised in Section~\ref{subsec:
  fixing times} that the regression bug fixing times have become slightly faster
recently.

\begin{table}[t!]
\centering
\caption{Coefficients in the Full Model}
\label{tab: full model}
\begin{tabular}{lrrr}
  \hline
 & Coef. & Std.~error & $p$-value \\
  \hline
(Intercept) & \textbf{4.58} & 1.44 & <0.01 \\
REPORTERS & 0.03 & 0.12 & 0.80 \\
BUGTRACKER: no (ref.) \\
BUGTRACKER: yes & 0.14 & 0.15 & 0.34 \\
REFERENCES & \textbf{0.07} & 0.01 & <0.01 \\
ENTROPY & -0.29 & 0.27 & 0.28 \\
REVTEST: no (ref.) \\
REVTEST: yes & -0.09 & 0.07 & 0.24 \\
ACKED: no (ref.) \\
ACKED: yes & -0.05 & 0.10 & 0.62 \\
SUGGESTED: no (ref.) \\
SUGGESTED: yes & -0.29 & 0.15 & 0.06 \\
MODFILE & -0.01 & 0.02 & 0.79 \\
LOCADD & \textbf{0.01} & <0.01 & <0.01 \\
LOCDEL & <0.01 & <0.01 & 0.11 \\
SUBSYSTEM: \texttt{drivers} (ref.) \\
SUBSYSTEM: \texttt{arch} & \textbf{0.46} & 0.12 & <0.01 \\
SUBSYSTEM: \texttt{block} & -0.50 & 0.32 & 0.12 \\
SUBSYSTEM: \texttt{crypto} & -0.30 & 0.76 & 0.69 \\
SUBSYSTEM: \texttt{fs} & -0.07 & 0.12 & 0.55 \\
SUBSYSTEM: \texttt{kernel} & \textbf{-0.34} & 0.17 & 0.04 \\
SUBSYSTEM: \texttt{lib} & \textbf{0.95} & 0.37 & 0.01 \\
SUBSYSTEM: \texttt{mm} & -0.05 & 0.21 & 0.82 \\
SUBSYSTEM: \texttt{net} & \textbf{-0.48} & 0.15 & <0.01 \\
SUBSYSTEM: \texttt{others} & \textbf{-0.41} & 0.16 & 0.01 \\
SUBSYSTEM: \texttt{security} & -0.32 & 0.76 & 0.68 \\
SUBSYSTEM: \texttt{sound} & -0.20 & 0.18 & 0.28 \\
Year 2021 (ref.) \\
Year 2022 & -0.13 & 0.18 & 0.48 \\
Year 2023 & -0.08 & 0.18 & 0.67 \\
Year 2024 & \textbf{-0.38} & 0.18 & 0.04 \\
\hline
\multicolumn{4}{l}{~~\small{Note: boldfaced coefficients denote $p < 0.05$.}}
\end{tabular}
\end{table}

\begin{table}[t!]
\centering
\caption{Performance of Consecutive Additive Models}
\label{tab: consecutive}
\begin{tabular}{lccc}
  \hline
Model\qquad\qquad\qquad\qquad & Parameters & Pseudo-R$^2$ & BIC \\
  \hline
Model 1. & 2 & <0.01 & 7938 \\
Model 2. & 3 & \phantom{<}0.06 & 7904 \\
Model 3. & 4 & \phantom{<}0.18 & 7825 \\
Model 4. & 5 & \phantom{<}0.18 & 7832 \\
Model 5. & 6 & \phantom{<}0.19 & 7837 \\
Model 6. & 7 & \phantom{<}0.19 & 7841 \\
Model 7. & 8 & \phantom{<}0.20 & 7842 \\
Model 8. & 9 & \phantom{<}0.20 & 7845 \\
Model 9. & 10 & \phantom{<}0.26 & \textbf{7811} \\
Model 10. & 11 & \phantom{<}0.26 & 7814 \\
Model 11. & 22 & \phantom{<}0.32 & 7839 \\
Model 12. & 25 & \phantom{<}\textbf{0.34} & 7845 \\
\hline
\multicolumn{4}{l}{~~\small{Note: boldfaced values denote a maximum and a minimum.}}
\end{tabular}
\end{table}

\begin{table}[th!b]
\centering
\caption{A Parsimonious Model Based on Automated Model Selection and Re-Coded SUBSYSTEM Dummy Variables}
\label{tab: model selection model}
\begin{tabular}{lrrr}
\hline
& Coef. & Std.~error & $p$-value \\
\hline
(Intercept) & \textbf{5.82} & 0.84 & <0.01 \\
REFERENCES & \textbf{0.07} & 0.01 & <0.01 \\
ENTROPY & \textbf{-0.56} & 0.16 & <0.01 \\
LOCADD & \textbf{0.01} & 0.00 & <0.01 \\
SUBSYSTEM\_R: \texttt{drivers} (ref.) \\
SUBSYSTEM\_R: \texttt{arch} & \textbf{0.50} & 0.12 & <0.01 \\
SUBSYSTEM\_R: \texttt{lib} & \textbf{1.10} & 0.38 & <0.01 \\
SUBSYSTEM\_R: \texttt{others} & \textbf{-0.25} & 0.08 & <0.01 \\
\hline
\multicolumn{4}{l}{~~\small{Note: boldfaced coefficients denote $p < 0.05$,}} \\
\multicolumn{4}{l}{~~~~~~~\small{Nagelkerke's pseudo-R$^2$ = $0.30$, and BIC = $7755$.}}
\end{tabular}
\end{table}

\begin{table}[th!b]
\centering
\caption{Classification Performance of the Full Model}
\label{tab: classification}
\begin{tabular}{lcccc}
\hline
& Precision & Recall & Accuracy & F1 \\
\hline
Na\"ive Bayes & 0.41 & \textbf{0.77} & \textbf{0.53} & 0.55 \\
Random forest & \textbf{0.45} & 0.60 & 0.51 & \textbf{0.58} \\
Support vector machine & 0.31 & 0.37 & 0.34 & 0.44 \\
\hline
\multicolumn{5}{l}{~~\small{Note: largest column-wise values are boldfaced.}} \\
\end{tabular}
\end{table}

Upon recalling Fig.~\ref{fig: revtest subsystem}, it can be also noted that
recomputing the full model with an additional interaction term between REVTEST
and SUBSYSTEM indicates that the \texttt{net} subsystem dummy variable is
statistically significant for the $\textmd{REVTEST} = 1$ interaction
case. Although the conclusion is thus not entirely definite regarding reviewing
and testing of commits fixing regression bugs, the main \texttt{net} subsystem
dummy variable can be reasonably taken to proxy also the subnormal reviewing and
testing within the subsystem.

With these points in mind, it seems sensible to expect that the full model could
be reduced to a more parsimonious one. To this end, Table~\ref{tab: consecutive}
shows the results from the manually computed sequential models that were
discussed in Section~\ref{subsec: methods}. The first thing to note from the
table is that the pseudo-R$^2$ value is $0.34$ for the full model. Although this
value indicates only modest overall statistical performance, it is by no means
atypical in the research branch~\cite{Ruohonen18IST}. Another point is that the
pseudo-R$^2$ values increase continuously as more parameters are added. However,
BIC, which penalizes more complex models heavily, indicates that LOCDEL,
SUBSYSTEM, and YEAR could be perhaps omitted for parsimony reasons.

The automated model selection algorithm yields even a more drastic observation:
interestingly enough, it only retains BUGTRACKER, REFERENCES, and LOCADD. This
algorithmic choice is likely because SUBSYSTEM contains too many dummy
variables, and hence it is being penalized heavily. For this reason, SUBSYSTEM
was re-coded by only retaining \texttt{drivers}, \texttt{arch}, and
\texttt{lib}, collating all other subsystems in to the category of
``others''. After recomputing the model selection algorithm with the re-coded
dummy variable set, the re-coded SUBSYSTEM is retained. The results are shown in
Table~\ref{tab: model selection model}. Although REFERENCES and LOCADD are again
present, their coefficients are small likely previously. In contrast, a
coefficient with a rather large magnitude and a negative sign is present for
ENTROPY. While it was noted in Section~\ref{subsec: metrics} that ENTROPY might
proxy interpretability of bug reports, the observation might instead prompt
theorizing that higher entropy implies more information for debugging, thus
shortening the bug fixing times, as indicated also by the coefficient's
sign. This line of reasoning would align with analogous arguments previously
raised in the literature~\cite{Ruohonen24SANERb}. However, the final important
point is that the parsimonious model does yield a BIC smaller than any of the
models in Table~\ref{tab: consecutive}, but its Nagelkerke's pseudo-R$^2$ value
is still lower than in the full model. As even the full model is relatively
small, the overall performance is modest, and there are no right or wrong
answers to a question whether parsimony should be preferred in these settings,
the full model in Table~\ref{tab: full model} can be taken as the preferred one
for answering to RQ.4's latter part. Thus, referencing emails and code churn
provide partial explanations, but stronger explanations are provided by the
subsystems in the kernel and the annual trend.

Finally, the classification approach provides a re-check regarding the overall
performance. The results are concisely summarized in Table~\ref{tab:
  classification}. The modest performance is again visible. In fact, the na\"ive
Bayes classifier, which interestingly outputs the highest accuracy value, as
well as the random forest classifier both indicate nearly a random
classification. While feature (that is, metric) selection might again change
things a little, it can be concluded that the regression bug fixing times cannot
be predicted well all in all. Computing the classifications with the consecutive
additive models does not bring any new relevant information; the performance is
generally better the higher the number of metrics included. Having said that, it
should be also emphasized that the split based on median is more or less
arbitrary~\cite{Ruohonen24SANERb}. When looking at the distribution in
Fig.~\ref{fig: diff time}, it would indeed seem more reasonable to try
multi-class classification, perhaps using quantiles as cutoff points. Similar
ideas have been presented also previously~\cite{Marks11}. However, additional
computing can be reasonably left for further work. In the meanwhile, the
negative binomial regression analysis provides a more elaborate conclusion.

\section{Discussion}\label{sec: discussion}

\subsection{Conclusion}

The answers to the five RQs can be summarized as follows:
\begin{itemize}
\itemsep 5pt
\item{Given a recent time period from 2021 to late 2024, the regression bug
  fixing times have been fast in the Linux kernel. The mean and median are only
  $25$ and $12$ days, respectively. The regression bug fixing times have also
  shortened recently.}
\item{Device drivers constitute the most prone subsystem for regression bugs in
  the Linux kernel. Over a half of the regression bugs observed have affected
  device drivers. Also the bug fixing times vary across the subsystems. When
  compared to device drivers, \texttt{arch} and \texttt{lib} have had slower
  bug fixing times, and some other subsystems faster fixing times.}
\item{The majority (60\%) of commits fixing regression bugs has been reviewed,
  tested, or both. However---with a notable exception of the \texttt{net}
  subsystem---the bug fixing times do not vary according to testing, reviewing,
  or both.}
\item{There is only a weak statistical signal that code churn might contribute
  to explaining the regression bug fixing times. Because the magnitude of the
  underlying effect is small and there are also outliers present, the overall
  conclusion is negative; code churn does not explain the bug fixing times well.}
\item{Although only modest overall statistical performance could be obtained for
  predicting the bug fixing times, several metrics show strong predictive
  effects. In addition to a decreasing annual trend, particularly the kernel's
  subsystems are relevant for modeling the fixing times. In addition referencing
  emails and code churn exhibit weak but still visible effects.}
\end{itemize}

In addition to these conclusions, the paper's remainder notes a few
limitations. After these notes, concluding remarks follow, including a few
points about the implications of the conclusions.

\subsection{Threats to Validity}

External validity---the extent to which results generalize to a wider
population---is a problem in any case study, including the present
work. However, the importance and relevance of the Linux kernel arguably make
the problem less severe. It should be also emphasized that the population of
operating system kernels has decreased over the decades; when a few commercial
operating systems are excluded, there are not many kernels with which a sample
could be increased. Nevertheless, the Berkeley software distributions (BSDs)
would provide interesting comparative cases for further work.

In quantitative empirical research construct validity refers to a question
whether the metrics used reflect what is intended to be measured. As was
discussed in~Section~\ref{subsec: metrics}, in the present work the associated
concern is not so much about the operationalization of the metrics but rather
about the difficulties in theorizing what they mean at a higher level of
abstraction. This theorizing---or lack thereof---should be acknowledged as a
problem. Having said that, it can be also argued that the bug fixing times
research branch as well as software metrics research in general have typically
been rather atheoretical. As will be soon discussed, addressing this lack of
theorizing would be a good topic for further research.

Internal validity generally refers to a study's execution, including its
statistical computations. In causal modeling, to which regression analysis also
belongs, appropriateness of an execution involves questions such as whether
confounding metrics were considered. Rather than such questions, however, the
overall modest statistical performance pinpoints toward the omitted variable
bias; some relevant information was missing from the analysis. The bias may also
involve the causal reasoning because the omitted information may not only affect
the bug fixing times but it may also correlate with the metrics used as
predictors~\cite{Wilms21}. Although this potential bias must be acknowledged as
an internal validity threat, addressing and fixing it would again provide a good
approach for further research.

\subsection{Further Work}

There are two paths along which further research might proceed: empirical and
theoretical. In terns of the former, a good starting point would be to examine
means by which the relatively modest overall statistical performance might be
improved. In addition to collecting more data, deriving and applying more
metrics might well improve statistical performance. For instance, the present
work did not consider the characteristics of bug reporters, committers, and
other developers, such as their expertise or reputation. Although usually only
modeled as latent features, such personal characteristics have been observed to
be good predictors for bug (and vulnerability) fixing
times~\cite{Ruohonen24SANERb}. Instead of considering and operationalizing
custom (direct) metrics, it would be also possible to focus on so-called
indirect measures~\cite{Kitchenham95}. Regardless of reporting venues, the bug
reports on which the bug fixing times are dependent are a good example in this
regard. For instance, it would be possible to use topic modeling for these
reports, assigning each report to a distinct topic, and then using the
assignments as predictors. Such an approach has been observed to increase
performance~\cite{Ruohonen17TIR}. However: when recalling the earlier points
about parsimony, the more there are metrics, the more difficult theorizing
is. Already for this reason, the theoretical path should be walked further
too. This path would also generally answer to the calls for better theoretical
engagement in software engineering research~\cite{Stol24}. In terms of the
present work, a starting point in theorizing would involve contemplating answers
to a simple question: what really explains the results reached?

The subsystems in the Linux kernel provide the most explanatory power in
statistical predictions. Thus, also an explanation for the regression bug fixing
times should at least partially come from these subsystems. The subsystems are a
core part of the overall software architecture of the Linux kernel. To this end,
extensive dependencies between subsystems may both increase a probability of
introducing regression bugs and lengthening regression bug fixing times. For
instance, a refactoring of a kernel interface may imply refactoring also
multiple device drivers due to their dependencies on the interface. A related
concept is scattering of features; instead of organizing kernel code in a
modular way, files and functionalities are extensively spread within a subsystem
or across multiple subsystems. Existing results indicate that Linux kernel
developers try to limit global scattering by making particularly device drivers
as self-contained as possible, but the device driver subsystem has still been
prone to scattering~\cite{Passos21}. Although these architectural reasons may
provide a good starting point for a theorization, it is also possible to
consider the subsystems from a social perspective.

Despite the dependencies and scattering, in the operating system context it can
be also argued that large subsystems form their own software architectures,
which are independent from a software architecture of an encompassing
system~\cite{Spinellis21}. This point can be extended toward the noted concept
of subsystem subcultures~\citep{Kudrjavets22a}. Although reviewing and manual
testing of regression bug fixes do not lengthen the associated fixing times
according to the results obtained, it may still be that subsystem-specific
software development practices, including social interactions and collaborations
between kernel developers, may contribute to explaining the fixing times.

To this end, previous results indicate that acceptance of patches and pull
requests does not depend only on technical merit but also social and strategic
aspects are often involved~\cite{Alami20}. Among the social aspects are trust
between people and mentoring of newcomers~\cite{Alami22}. Such social aspects
may well provide additional explanatory power. Social aspects are important also
in terms of human resources required to develop and maintain the Linux
kernel. Existing results indicate that most newcomers are working with device
drivers, whereas the amounts of developers working in some other subsystems are
stagnating or even declining~\cite{ZhouCheng17}. The traditionally well-balanced
ratio of contributors to reviewers in the Linux kernel~\cite{Bettenburg15} may
have also changed in some particular subsystems. Thus, potentially increasing
workloads in some subsystems may well contribute to lengthening of regression
bug fixing times in these subsystems. A final point about human resources
relates to the Linus law: it may be that users report more regression bugs in
device drivers because these are often visible to them in a sense that a device
may no longer work or the whole kernel even crashes. To this end, the slow
fixing times in the \texttt{lib} subsystem may be because only a few people are
looking at the code therein. To proceed with these points further, a more
thorough theorization is required.

Finally, it is easy to agree with previous studies in that the Linux kernel
would benefit from more regression testing~\cite{Xiao19}. Though, that might be
easier said than done because testing may require real hardware. Device drivers
are again a good example in this regard.

\balance
\bibliographystyle{abbrv}

\end{document}